\documentclass[floatfix,aps,twocolumn,showpacs,superscriptaddress,amsmath,amssymb]{revtex4-1}

\usepackage[autostyle=true]{csquotes}

\usepackage{mathtools}
\usepackage{textcomp}
\usepackage{gensymb}
\usepackage{graphicx}
\usepackage{siunitx}
\usepackage{ulem}

\usepackage{color}

\usepackage[unicode=true,
 bookmarks=true,bookmarksnumbered=true,bookmarksopen=true,bookmarksopenlevel=2,
 breaklinks=false,pdfborder={0 0 1},backref=false,colorlinks=true]
{hyperref}

\hypersetup{linkcolor=blue, citecolor=blue, urlcolor=blue, filecolor=blue, pdfpagelayout=OneColumn,
  pdfnewwindow=true, pdfstartview=XYZ, plainpages=false}

\usepackage[all]{hypcap}

\begin{document}

\def\be{\begin{equation}}
\def\ee{\end{equation}}
\def\ki{\kappa_i}
\def\kf{\kappa_f}
\def\kopt{\kappa_{opt}}
\def\tr{\tau_{relax}}
\def\pN{pN/$\mu$m}

\title{Noise and ergodic properties of Brownian motion in an optical tweezer: looking at the crossover between Wiener and Ornstein-Uhlenbeck processes}

\author{R\'emi Goerlich}
\email{Equal contributions}
\affiliation{Universit\'e de Strasbourg, CNRS, Institut de Physique et Chimie des Mat\'eriaux de Strasbourg, UMR 7504, F-67000 Strasbourg, France}
\affiliation{Universit\'e de Strasbourg, CNRS, Institut de Science et d'Ing\'enierie Supramol\'eculaires, UMR 7006, F-67000 Strasbourg, France}
\author{Minghao Li}
\email{Equal contributions}
\affiliation{Universit\'e de Strasbourg, CNRS, Institut de Science et d'Ing\'enierie Supramol\'eculaires, UMR 7006, F-67000 Strasbourg, France}
\author{Samuel Albert}
\affiliation{Universit\'e de Strasbourg, CNRS, Institut de Science et d'Ing\'enierie Supramol\'eculaires, UMR 7006, F-67000 Strasbourg, France}
\author{Giovanni Manfredi}
\email{giovanni.manfredi@ipcms.unistra.fr}
\affiliation{Universit\'e de Strasbourg, CNRS, Institut de Physique et Chimie des Mat\'eriaux de Strasbourg, UMR 7504, F-67000 Strasbourg, France}
\author{Paul-Antoine Hervieux}
\email{hervieux@ipcms.unistra.fr}
\affiliation{Universit\'e de Strasbourg, CNRS, Institut de Physique et Chimie des Mat\'eriaux de Strasbourg, UMR 7504, F-67000 Strasbourg, France}
\author{Cyriaque Genet}
\email{genet@unistra.fr}
\affiliation{Universit\'e de Strasbourg, CNRS, Institut de Science et d'Ing\'enierie Supramol\'eculaires, UMR 7006, F-67000 Strasbourg, France}

\date{\today} 
\begin{abstract}

We characterize throughout the spectral range of an optical trap the nature of the noise at play and the ergodic properties of the corresponding Brownian motion of an overdamped trapped single microsphere, comparing experimental, analytical and simulated data. We carefully analyze noise and ergodic properties $(i)$ using the Allan variance for characterizing the noise and $(ii)$ exploiting a test of ergodicity tailored for experiments done over finite times. We derive these two observables in the low-frequency Ornstein-Uhlenbeck trapped-diffusion regime and study analytically their evolution towards the high-frequency Wiener free-diffusion regime, in a very good agreement with simulated and experimental results. This leads to reveal noise and ergodic spectral signatures associated with the distinctive features of both regimes.

\end{abstract}

\maketitle

\section{Introduction}

The high sensitivity of optically trapped Brownian particles, combined with long integration times available, makes optical traps outstanding metrological systems.
They have therefore been involved in many weak force experiments and have been recognized as outstanding systems for implementing and simulating many results and protocols that have been brought forward recently in the field of optomechanics and non-equilibrium statistical physics \cite{CilibertoPRX2017,Martinez2017,Bechhoefer_2020}.

Optical traps physically implement an Ornstein-Uhlenbeck process through the harmonic trapping force field. One of their interesting features is to give access to different diffusing dynamics for the trapped Brownian object, ranging from confined motion in the long timescales to free Brownian motion on the shortest ones, therefore probing the evolution of the Ornstein-Uhlenbeck process towards the Wiener process-like limit at short times  \cite{Uhlenbeck,gardiner}. These two dynamic regimes have very different properties that make them more relevant for different experiments. In particular, the Ornstein-Uhlenbeck regime is well suited for force measurements \cite{Wu2009,MaiaNeto2015,Ricci2017,Li} while position detection benefits from the Wiener regime, allowing to achieve higher resolution \cite{Lukic2005,li2010,Schnoering2019}.

In this article, we address these differences from the viewpoints of noise stability and ergodicity for both regimes. We implement theoretical and experimental tools capable of characterizing motional noise (using an Allan-variance based analysis \cite{Oddershede,Saleh}) and ergodic signatures (developing a specific test of ergodicity \cite{Metzler2014,Metzler2015}) in an optical trap from the Ornstein-Uhlenbeck regime to its high frequency Wiener limit in a unified way.

This capacity is important in particular in the field of precision measurements involving optical traps. There indeed, the building of large motional statistical ensembles necessary to reach high resolution levels usually relies on strong assumptions related to the nature and stability of the driving noise. It also depends on the ergodicity of the corresponding Brownian motion. We show here precisely how these assumptions can be tested on overdamped harmonic optical traps, paving the way for reliable experiments at all measurement bandwidths. The tools we describe below are general: they can be used on underdamped and more complex systems and can thereby be exploited when colored noise and non-ergodic effects enrich the physics of Brownian motion, as in the realm of swimmers or active matter, for instance.

\section{Wiener vs. Ornstein-Uhlenbeck crossover in an optical trap}

Free Brownian motion driven only by the Gaussian white noise of thermal fluctuations is described by the Wiener process $W_t$. The displacement of the overdamped free Brownian object writes as:
\begin{equation}
	dx_t = \sqrt{2D}dW_t  , 
	\label{wiener}
\end{equation}
working directly with the differential $dW_t$ with the following properties: $\langle dW_t \rangle = 0$, $\langle dW_t dW_{t'} \rangle = \delta(t, t') dt$.The diffusion coefficient $D = k_B T / \gamma$ involves the Boltzmann constant $k_B$, the temperature of the surrounding fluid $T$ and the Stokes drag coefficient $\gamma$.

Inside the trap, the harmonic optical potential modifies the stochastic process by exerting on the object a restoring force characterized by a constant stiffness $\kappa$. The same displacement now follows the Ornstein-Uhlenbeck process:
\begin{equation}
	dx_t = -\frac{\kappa}{\gamma}x_tdt + \sqrt{2 D}dW_t .
	\label{ou}
\end{equation}

\begin{figure}[htb]
  \centering{
    \includegraphics[width=0.8\linewidth]{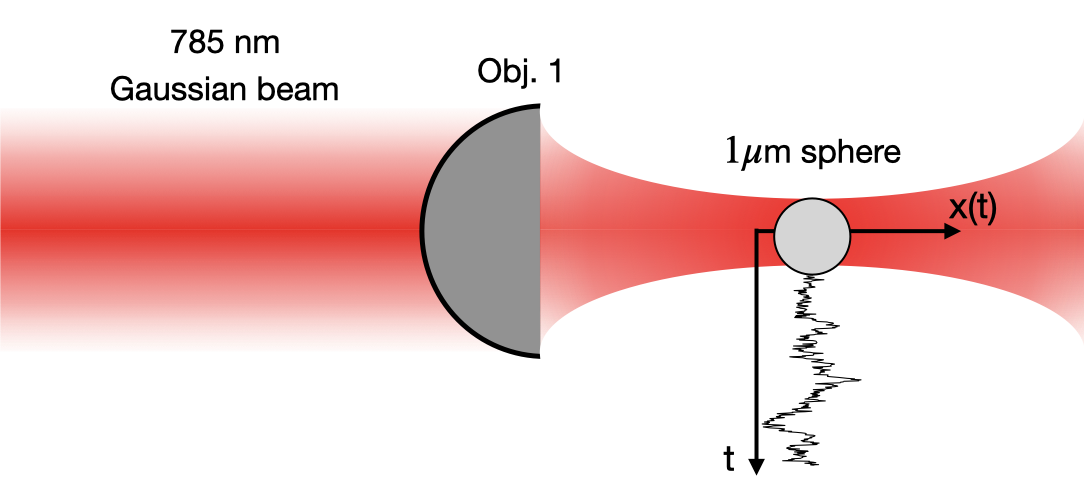}}
  \caption{{Schematic view of the optical trapping system: a $1~\mu$m polystyrene bead is trapped by a 785 nm laser beam, focused by a high numerical aperture (NA=$1.2$) water immersion objective. The instantaneous position of the bead trapped at the laser waist is recorder along the optical axis with an acquisition frequency of $2^{15}=32768$ Hz.}}
  \label{schema}
\end{figure}
\vspace{3mm}

Our experiment, detailed in Appendix \ref{APPENDIX_samples}, consists in trapping a single Brownian object in the harmonic potential created at the waist of a focused laser beam, and recording the instantaneous overdamped position $x(t)$ of the trapped bead, as schematized in Fig. \ref{schema}. All the experimental results presented in this work are obtained from a 10 minutes long trajectory (i.e. $1.97\times 10^{7}$ successive position measurements acquired at a frequency of $2^{15}=32768$ Hz).

These data are compared, throughout this article, with numerical simulations obtained from an algorithm for the Wiener process:
\begin{equation}
	x_{t+\Delta t} = x_t + \sqrt{2D\Delta t} \theta_t ,
\end{equation}
where $\theta$ is a dimensionless Gaussian white noise with $\langle \theta_t \rangle = 0$, $\langle \theta_t \theta_{t'} \rangle = \delta(t-t')$, according to the methods detailed in \cite{volpe}. By the same token, the algorithm for the Ornstein-Uhlenbeck process is:
\begin{equation}
	x_{t+\Delta t} = x_t - \frac{\kappa}{\gamma}x_t \Delta t + \sqrt{2D\Delta t} \theta_t .
\end{equation}
This discretisation method, known as the Euler-Maruyama method, corresponds to an $ \mathcal{O}(\Delta t^{1/2})$ approximation of It\^o-Taylor expansions \cite{kloeden}. As discussed in details in Appendix \ref{APPENDIX_algorithm}, higher order terms lead to a more efficient algorithm known as the Mildstein algorithm, which our simulations are based on and which converges more quickly towards the analytical expression as $\Delta t$ decreases \cite{Higham2001,vanden-eijnden}. 

\begin{figure}[htb]
  \centering{
    \includegraphics[width=1\linewidth]{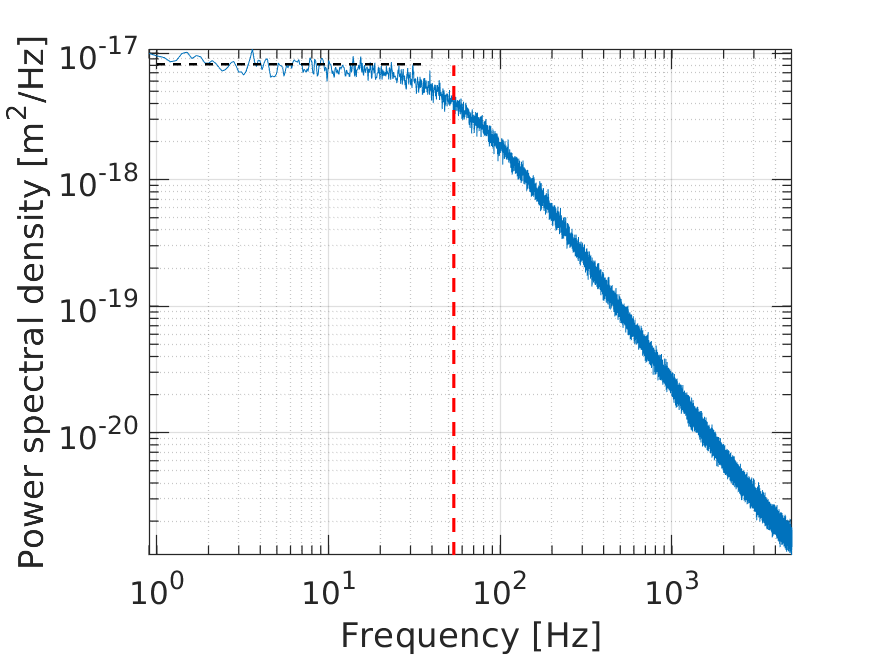}}
  \caption{{Experimental power spectrum density (PSD) evaluated for a trajectory $x(t)$ measured from $0.03$ \si{\hertz} to $100$ \si{\kilo \hertz}, displaying a large signal-to-noise ratio, spanning over 4 decades. We also note the transition, at the roll-off frequency ($53.6511$ Hz) between the high frequency almost-free regime and the low frequency trapped regime -vertical red dashed line. The thermal noise plateau $2 k_B T \gamma / \kappa $ (horizontal black dashed line) agrees well with the low frequency limit of the PSD, as expected. From the Lorentzian fit performed on the PSD, we can extract the stiffness $\kappa = 2.9614\cdot10^{-6} \pm 6.7339\cdot10^{-8}$ kg/s$^2$. The experiments are performed at room temperature, $T \approx 295$ \si{\kelvin} and the 1 \si{\micro\meter} bead experiences a drag coefficient $\gamma = 6 \pi \eta R $ kg/s where $\eta \approx 0.95 \cdot10^{-3}$ \si{\pascal\second}, hence $\gamma = 8.9837\cdot 10^{-9}$ kg/s. These parameters, with the stiffness extracted from the Lorentzian fit of the PSD, are used in all numerical and analytical results done throughout the paper.}}
  \label{psd}
\end{figure}

From Eq. (\ref{ou}), the Brownian motion in the trap can be spectrally analyzed with the position's power spectral density (PSD):
\begin{equation}
 	S_{x}(f) = \frac{D}{2\pi^2(f_c^2 + f^2)} .
 	\label{psd_eq}
 \end{equation}
As clearly seen on the experimental PSD displayed in Fig. \ref{psd}, the roll-off frequency $f_c = \kappa/ (2 \pi \gamma)$ separates the high frequency regime $S_x(f)\sim{D}/{(2\pi^2f^2)}$ of free Brownian motion -see Eq. (\ref{wiener})- from the low frequency trapping regime $S_{x}(f) \sim {D}/{( 2 \pi^2 f_c^2)} = 2 k_B T \gamma / (\kappa^2) $ -see Eq. (\ref{ou}). The PSD thus clearly reveals how a Wiener regime corresponds in the optical trap to the short time $\delta t \ll \gamma/\kappa$ limit of the Ornstein-Uhlenbeck process (in other words, when observed over such a short timescale, the Brownian object moves inside the trap as if it were freely diffusing without confinement). 

\vspace{3mm}

\section{Noise stability: Allan variance and statistical tests}  

In order to characterize the noise at play inside the optical trap, it is central to measure two of its properties: its nature (color, thermal weight, frequency contributions, etc), and its stability in time. Testing the nature of the noise can be done spectrally with the PSD that yields the different frequency contributions of the noise. Integrated PSD can also reveal the thermal nature of the noise through the fluctuation-dissipation theorem. However, the spectral approach turns out to be exposed to possible low frequency drifts that can modify noise properties \cite{Li,Oddershede,Saleh}. In order to avoid this stability issue, we work in the time-domain and perform an Allan-variance based test of the system, capable of revealing low frequency drifts within a stochastic signal \cite{Allan1966,Barnes1971}. This approach leads us to verify unambiguously the stationary and thermally limited properties of the noise at play in an experiment.

The Allan variance $\sigma^2(\tau)$ can be connected to the noise PSD $S(f)$ through the following relation \cite{Barnes1971}:
\begin{equation}
	\sigma^2(\tau) = \frac{4}{\pi \tau^2} \int_{-\infty}^{+\infty} S(f)\sin^4(\pi f \tau) df
\end{equation}
It can therefore be explicitly evaluated analytically for the Ornstein-Uhlenbeck PSD $S_x(f)$ of Eq. (\ref{psd_eq}):
\begin{equation}
	\sigma^2(\tau) = \frac{k_B T}{\kappa \tau^2} \left( 4\left[1 - e^{-\kappa\tau/\gamma}\right] - \left[1 - e^{-2\kappa\tau/\gamma}\right]\right)   ,
	\label{allanvar_eq}
\end{equation}
as detailed in Appendix \ref{APPENDIX_allan}.

\begin{figure}[htb]
  \centering{
    \includegraphics[width=1\linewidth]{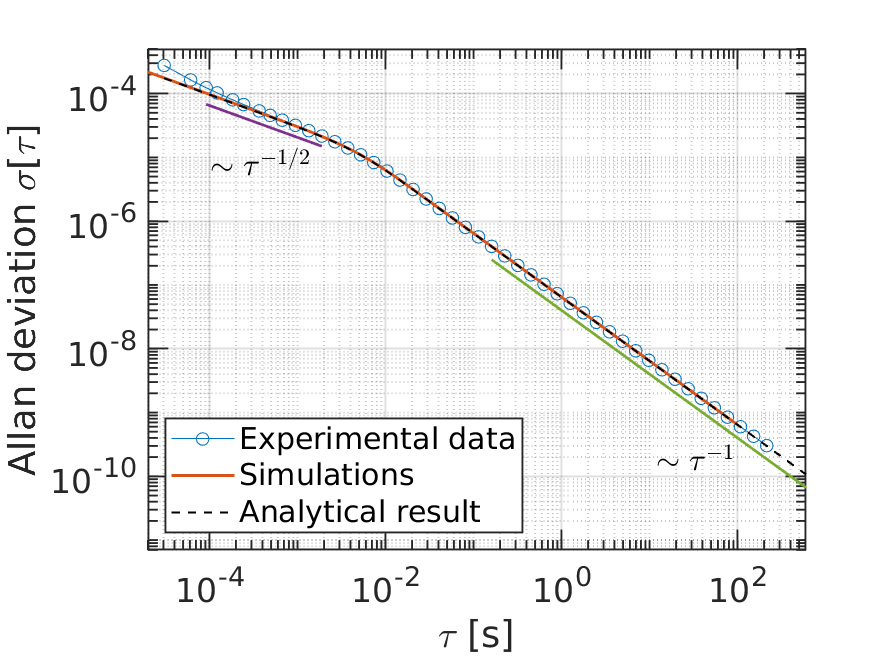}}
  \caption{{Allan standard deviation evaluated for the long trajectory experimentally recorded (blue open circles). We plot the simulated Allan standard deviation (orange continuous line) superimposed to the analytical result (black dashed line). We highlight the slopes in both free (purple continuous line) and trapped regimes (green continuous line). We observe that the whole time range from $\sim 10^{-4}s$ up to $\sim10^{2}s$ is perfectly captured by the theoretical expression built with experimental parameters --$\gamma, T, \kappa$, see Fig. \ref{psd}-- with a very good agreement. The small departure of the experimental data from the theoretical Allan variance is attributed to tracking errors discussed in Appendix \ref{APPENDIX_err_allan}.}}
  \label{allan}
\end{figure}

The experimental Allan variance is shown in Fig. \ref{allan} following the same methodology presented in our earlier work \cite{Li}. This experimental Allan variance is compared with numerical simulations and with the analytical result of Eq. (\ref{allanvar_eq}). We note a remarkable experiment-theory agreement over more than 6 decades in time. These results show the very high level of noise stability up to $>250$ s that one can reach on a simple optical trap setup such as ours. 

But they also reveal how the Ornstein-Uhlenbeck and the Wiener processes are characterized by different Allan variance signatures. Indeed, we identify here two clear asymptotic regimes. The short time regime  ($\tau \ll \gamma/\kappa$) falls on the $\sigma_{\rm free} \sim t^{-1/2}$ slope, which is known to corresponds to the thermal white noise limit of free Brownian motion \cite{Oddershede,Li}. Interestingly, in the long time limit ($\tau \gg \gamma/\kappa$) of the Ornstein-Uhlenbeck process where the trapping action dominates the motional dynamics, the Allan variance shows a different slope with $\sigma_{trap} \sim t^{-1}$. 
This change of signatures between the two regimes, accounting for the presence of the harmonic force field in the long time limit, is continuous. We observe a very good match between the experiments and theory in the transition between asymptotic regimes.

The slight differences at short time-lags between the theory and the experimental data will also be observed at the level of the mean squared displacement (MSD) Fig. \ref{MSD_fig} (a) and the ergodic parameter Fig. \ref{ergodicity}. As discussed in details in Appendix \ref{APPENDIX_err}, these deviations are due to tracking errors unavoidably induced experimentally by the photodiode and electronic system used for recording our Brownian trajectories.

\vspace{3mm}

We will now use an alternative method based on the autocorrelation and the MSD for identifying either a Wiener or an Ornstein-Uhlenbeck process. We however remind here that at thermal equilibrium, Wiener and Ornstein-Uhlenbeck processes generate trajectories $x(t)$ with different statistical properties. Indeed, the Ornstein-Uhlenbeck process of the trapped Brownian motion has a variance constant in time with the equipartition condition $\langle x_t^2 \rangle = k_B T / \kappa$.
In contrast, the Wiener process of free Brownian motion is non-stationary with a motional variance that grows linearly in time.
But looking at the statistical properties of successive displacements $dx_t$ whose dynamics is governed by Eqs. (\ref{wiener},\ref{ou}), it becomes possible to perform the same stationarity test for both processes. To do that, we will use the autocorrelation of displacements and the MSD, extracted from long trajectories. 
We will verify stationarity --in the strong sense since the noise is Gaussian-- with $(i)$ a fixed mean (that can be removed without any loss of generality), $(ii)$ a finite variance $\overline{dx_t^2} $, and $(iii)$ a displacement covariance (autocorrelation) $\overline{dx_t dx_s}$ that depends only on the absolute time difference $\Delta = \mid t - s \mid$. 

The covariance of displacements can be computed using Eq. (\ref{ou}) (details are given in Appendix \ref{APPENDIX_cov}, see Eq. (\ref{autocorrdim})) and yields:
\begin{equation}
		\overline{dx_t dx_s} = -\frac{2\kappa k_B T}{\gamma^2} e^{-\kappa|t-s|/ \gamma} dt^2 + 2 D \delta(t-s) dt.
	\label{cov}
\end{equation}
This theoretical expression is compared to the covariance evaluated experimentally as a time-average on successive displacements. The comparison, together with simulations, is shown on figure \ref{MSD_fig} (a). The convergence of the time averaging process for the covariance towards the theoretical expression, only function of $\Delta$, shows the absence of dependence on the absolute time $t$. 

We can also evaluate the MSD directly from the measurement of successive positions separated by a given time-lag $\Delta$ (details are given in Appendix \ref{APPENDIX_MSD}, see Eq. (\ref{MSD_analytic})) as:
\begin{equation}	 
	\overline{\delta x^2(\Delta) } = 2 \frac{k_B T}{\kappa}\left(1 - e^{-\kappa \Delta/\gamma}\right) .
	\label{msd}
\end{equation}
Again, this theoretical result is compared to the experimental MSD which is given by evaluating the time average MSD of the entire trajectory. The comparison, also including simulations, shows a very good agreement displayed in Fig. \ref{MSD_fig} (b). 
 
This agreement, together with the covariance, depending only on time-difference, confirms that our Brownian trap implements a strong stationary Ornstein-Uhlenbeck process. Clearly, our data demonstrate a smooth crossover between the linear MSD at short time lags associated with a Wiener regime and the constant MSD at longer time lags that reflects the confined nature of the diffusion for the Ornstein-Uhlenbeck process.

\begin{figure}[htb]
  \centering{
    \includegraphics[width=0.9\linewidth]{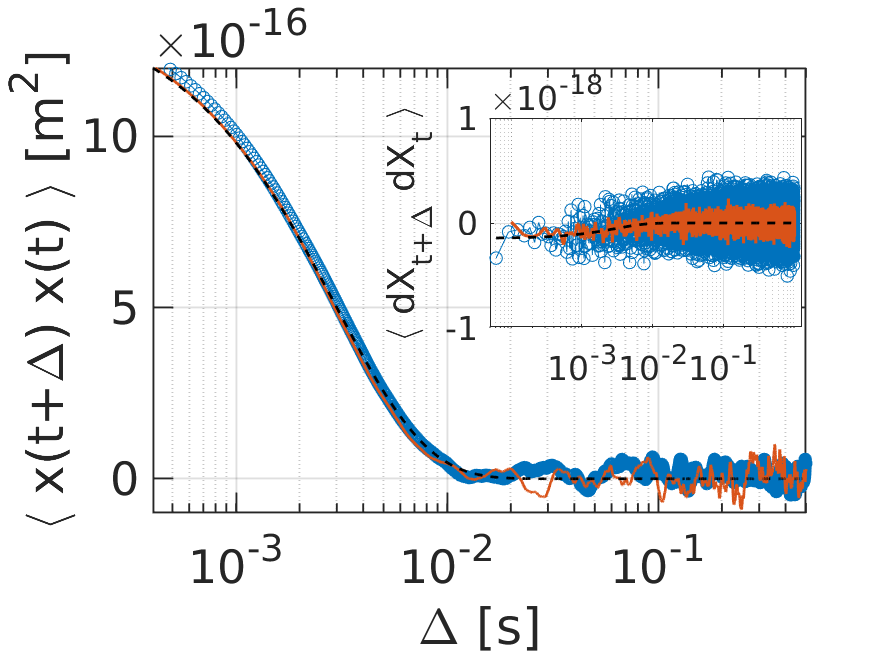}
    \includegraphics[width=0.9\linewidth]{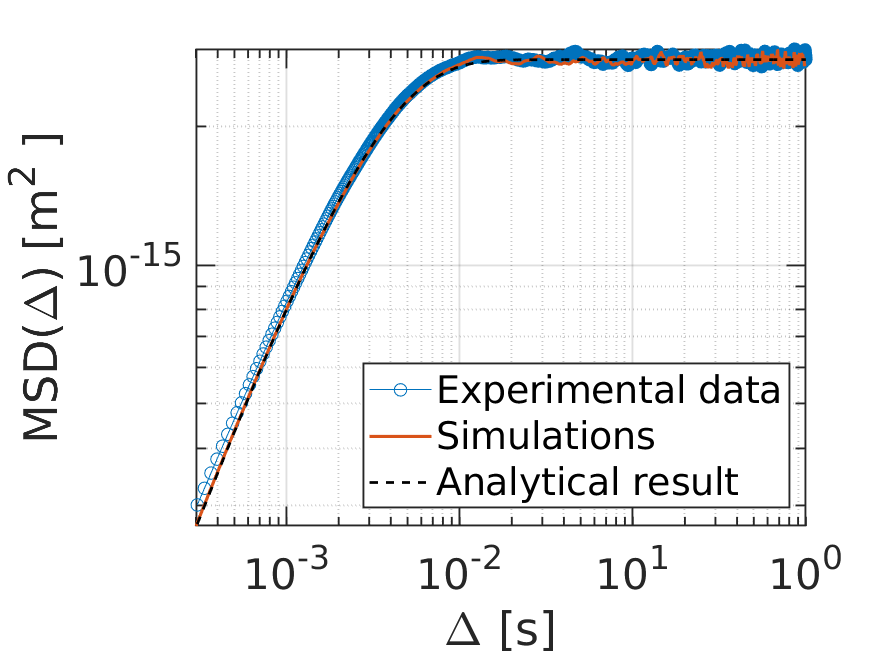}
    }
  \caption{{(a) Time average covariances of positions and displacements (inset). Experimental data are plotted (blue open circles) together with the simulation results (orange continuous line) and the analytical prediction (black dashed line). (b) Comparison between the measured mean square displacements (MSD) (blue open circles) and the analytical expression given in Eq. (\ref{msd}) obtained in the stationary regime (black dashed line). The comparison with simulation results is also displayed (orange continuous line). The very good agreement with both theory and simulations shows that the measured process can be considered as stationary. We note the same relaxation time of $ 3 \cdot 10^{-3} $s for all data, revealing the crossover between the free (Wiener) and trapped (Ornstein-Uhlenbeck) diffusion regimes. Again, the small departure of the experimental data with respect to the theoretical MSD is attributed to tracking errors discussed in Appendix \ref{APPENDIX_err_msd}.}}
  \label{MSD_fig}
\end{figure}

\section{Test of ergodicity}
\label{SEC:ergo}

As reminded in the Introduction, the ergodic hypothesis is central for reaching high resolution levels in optical trapping experiments. Ergodicity \textit{per se} corresponds to the equality taken in the infinite time limit $\mathcal{T} \rightarrow \infty$, between the time average and the ensemble average for a given stochastic process. In order to test ergodicity, we first need to build an ensemble of trajectories $\{i\}$. To do this, we reshape our long trajectory into an ensemble of 600 trajectories $x_i(t)$ of 1 second duration each. 
For such a trajectory $x_i(t)$ drawn from the ensemble, ergodicity is defined as:
\begin{equation}
\lim\limits_{\mathcal{T} \rightarrow \infty} \frac{1}{\mathcal{T}} \int_0^{\mathcal{T}} x_i(t)dt = \langle x_i(t) \rangle_{\{ i\}} .
\end{equation}

Although simple, this definition is however hardly operative in experiments that only yield ensembles of finite-time trajectories. Following the approach proposed in \cite{Metzler2014,Metzler2015}, we prefer resorting to an observable that can characterize the ergodic nature of an experiment performed over a finite integration time. This observable is grounded on the stationary nature of the MSD which is, as we shown above, independent of the choice from the initial time and only depends on the time-lag $\Delta$. In such conditions, ergodicity simply demands the time average MSD of any $i^{\rm th}$-trajectory, as defined above, to be equal, in the long $\mathcal{T}/\Delta$ limit, to the ensemble mean of individual time average taken over the ensemble $\{i\}$ of available trajectories:
\begin{equation}
\lim\limits_{\mathcal{T}/\Delta \rightarrow \infty}  \overline{\delta x_i^2(\Delta)}= \biggl<  \overline{\delta x_i^2(\Delta)} \biggl> .
\end{equation}

Formally, ergodicity demands that the $ \overline{\delta x_i^2(\Delta)}/ \biggl<  \overline{\delta x_i^2(\Delta)} \biggl>$ ratio tends to a Dirac distribution as $\mathcal{T}/\Delta \rightarrow \infty$. A sufficient condition for ergodicity is therefore that the normalized variance of this ratio goes to zero in the limit $\mathcal{T}/\Delta \rightarrow \infty$:
\begin{equation}
\epsilon(\Delta) = \frac{\biggl<  \overline{\delta x_i^2(\Delta)}^2 \biggl> - \biggl<  \overline{\delta x_i^2(\Delta)} \biggl>^2}{\biggl<  \overline{\delta x_i^2(\Delta)} \biggl>^2} .
\end{equation}

\begin{figure}[htb]
  \centering{
    \includegraphics[width=1\linewidth]{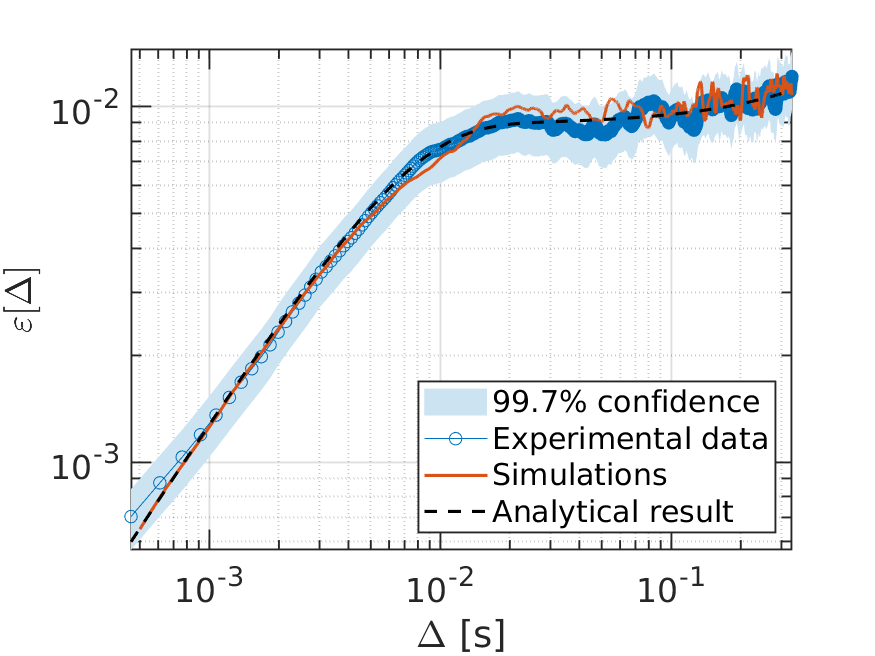}}
  \caption{{The normalized variance $\epsilon(\Delta)$ playing the role of an ergodic parameter is displayed (black dashed line) when calculated for the Ornstein-Uhlenbeck process at play in our optical trap. Experimental results (bleu open circles) for $\epsilon(\Delta)$ are compared to the theory within a $99.7\%$ confidence interval. We also show the results of a numerical simulation using $\mathcal{O}(3/2)$ algorithm (orange continuous line). The slight deviation at short times between the experiment and the theory comes again mainly from the position tracking errors whose impact on the ergodic parameter is discussed in Appendix \ref{APPENDIX_err_erg}.}}
  \label{ergodicity}
\end{figure}
\vspace{3mm}

Handling therefore finite integration times, this normalized variance $\epsilon(\Delta)$ is the right observable needed to prove the ergodic nature of a stochastic process experimentally implemented. One very appealing aspect of $\epsilon(\Delta)$ is that it can be theoretically calculated for an Ornstein-Uhlenbeck process, as we do in Appendix  \ref{APPENDIX_ergo}. This gives the capacity to characterize the ergodicity throughout the spectral range of the optical trap, therefore both in the long-time trapped and the short-time free diffusion regimes. These two regimes correspond to different time-lag evolutions of $\epsilon(\Delta)$, as clearly seen in Fig. \ref{ergodicity}. Here too, a smooth crossover between the long time-lag trapped (Ornstein-Uhlenbeck) regime and the short time-lag free (Wiener limit) regimes is revealed and measured, with the transition time-lag determined from the trap stiffness, as discussed in more details in Appendix \ref{APPENDIX_ergo}. The experimental evolution of $\epsilon(\Delta)$ corresponding to the recorded finite-time trajectories obtained for our trapping experiment is also shown. The excellent agreement with the theoretical $\epsilon(\Delta)$ in both the freely diffusing and in the trapped regimes confirms that our optical trapping process can be considered as ergodic with a high level of confidence. Because $\epsilon(\Delta)$ is formally a variance, the quality of its estimator on a finite-size ensemble can be quantified using a $\chi^2$-test. We perform in Fig. \ref{ergodicity} this test up to a $3\sigma$ level of confidence. 

\section{Conclusion}

By implementing in a combined manner Allan variance-based, stationarity and ergodic tests, we have been able to fully characterize, through wide spectral ranges, the nature of the noise and the ergodicity of the stochastic regimes at play in our overdamped optical trap.
In particular, our observables have revealed distinctive features between the high and low frequency range of the trap.
There are clear differences from the viewpoint of noise stability and ergodicity between Wiener and Ornstein-Uhlenbeck processes notwithstanding that they are driven by the same Gaussian white thermal noise.
These differences appear in our results when comparing the different dynamical regimes.
In stochastic thermodynamics, ergodic processes are a very important subclass of stationary processes. When aiming at exploiting Brownian systems, it is therefore very important to be able to identify stationarity signatures.  The simple and straightforward methodology proposed in our work is also relevant to many recent experiments involving Brownian systems coupled to non-thermal, colored, and more complex noise environments 
\cite{VolpeRMP2016}.

\section{Acknowledgements}

Thanks are due to A. Canaguier-Durand and G. Schnoering for stimulating discussions. This work was supported in part by Agence Nationale de la Recherche (ANR), France, ANR Equipex Union (Grant No. ANR-10-EQPX-52-01), the Labex NIE projects (Grant No. ANR-11-LABX-0058-NIE), and USIAS within the Investissements d'Avenir program (Grant No. ANR-10-IDEX- 0002-02).


%
\appendix

\section{Experimental setup}
\label{APPENDIX_samples}

\begin{figure}[htb]
  \centering{
    \includegraphics[width=1\linewidth]{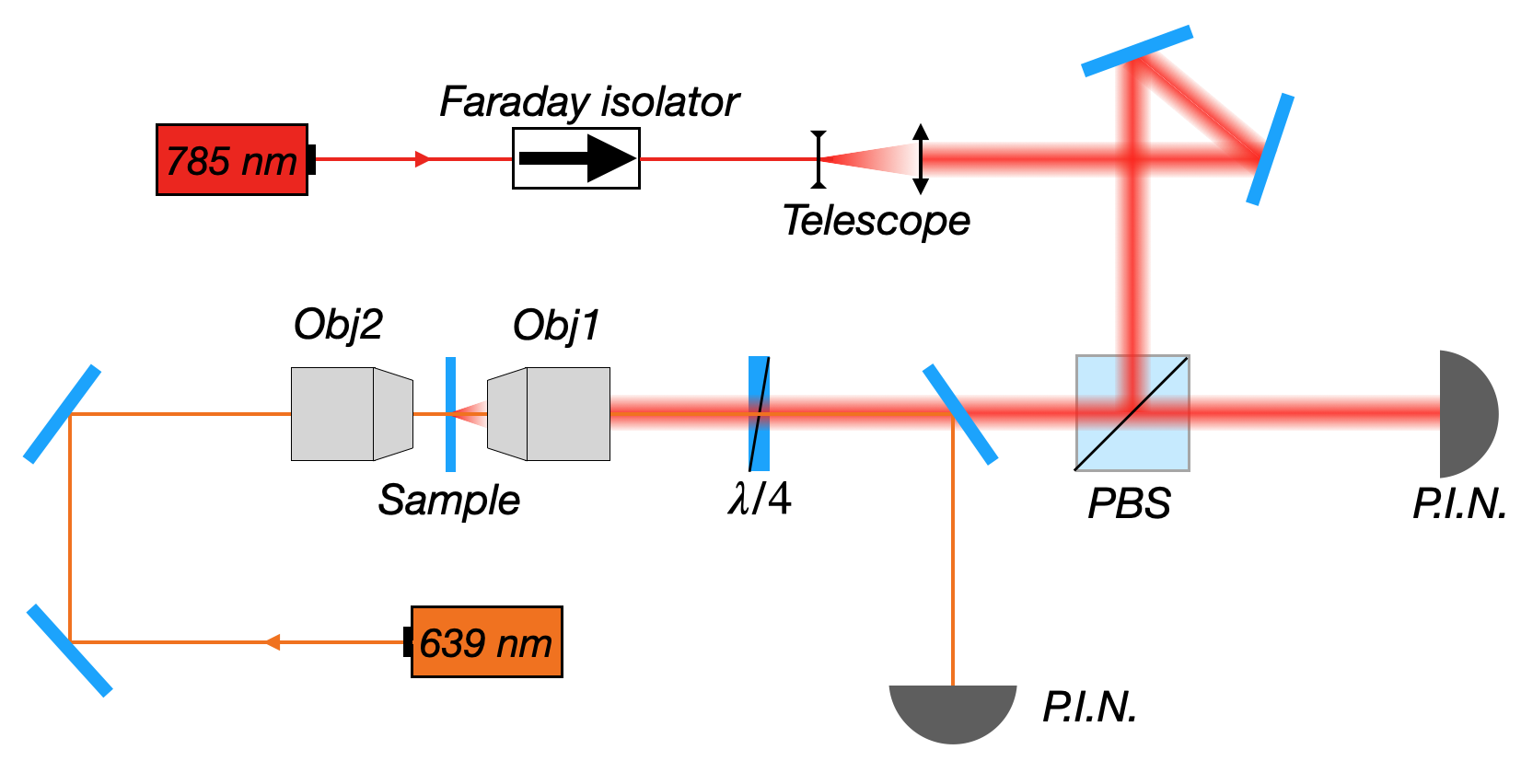}}
  \caption{{Schematic view of the main elements of the experimental setup, displaying the trapping laser (785 nm) and the laser (639 nm) used for recording instantaneous displacement of the trapped sphere. The trapping laser is sent to the sample using a polarizing beam splitter (PBS) and a water immersion objective (Obj1). The quarter wave-plate ($\lambda / 4$) ensures that the light scattered by the trapped bead and collected by Obj1 is directed towards the on-axis P.I.N. photodiode. The probe laser illuminates the bead from the backside using a second objective (Obj2) and is collected by Obj1. A dichroic beam-splitter sends the probe light to the second P.I.N. photodiode.}}
  \label{APPENDIX:schema}
\end{figure}

Our experiment consists in trapping a single Brownian object in the harmonic potential created at the waist of a focused laser beam. A schematic view of the setup is given on Fig.  \ref{APPENDIX:schema}. A linearly polarized Gaussian beam (OBIS Coherent, CW 785 nm, 110mW) is focused by a water immersion objective (Nikon Plan Apochromat $60\times$, Numerical Aperture $1.20$) into the sample that consists in a cell made of a glass slide and a coverslip, separated by a $120~\mu$m  thick and $1$ cm wide spacer. The cell is filled with a colloidal dispersion of polystyrene microspheres (ThermoFisher FluoSpheres polystyrene microspheres, $1~\mu$m diameter $\pm 2\% $) diluted in deionised water. We start with a solution of concentration of $10^{10}$ beads/mL that we dilute $\sim 10^5 \times$. The cell is then taped to a metallic holder mounted in our optical setup.

\vspace{3mm}

The instantaneous position of the trapped single bead is recorded using an additional low-intensity counter-propagating laser beam (Thorlabs HL6323MG CW $639$ nm, 30 mW, but here used at low power), focused on the bead using a second objective (Nikon Plan-fluo Extra Large Working Distance $60\times$, Numerical Aperture $0.7$). Within the small trapping volume defined by our setup, the intensity of the light scattered by the microsphere scales linearly with its displacement $x(t)$ along the optical axis. This scattered intensity signal is collected through the trapping objective and sent to a P.I.N. photodiode (Thorlabs, model Det10A2). The output signal recorded in V is sent to a low noise amplifier (Stanford Research, SR560) and then acquired by an analog-to-digital card (National Instrument, PCI-6251). The signal is filtered through a $0.3$ Hz high-pass filter at 6 dB/oct in order to remove the DC component of the output signal and through a $100$ kHz low-pas filter at 6 dB/oct to prevent aliasing. Finally, we convert the voltage signal into displacements measured in m.

In our experiment, it is crucial to trap only one bead at a time. To achieve this, we rely on $(i)$ a low concentration of beads in the solution and $(ii)$ a direct imaging of the vicinity of the trap with an Interferometric scattering microscopy technique (not shown on the figure but presented in details in our previous work \cite{SchnoeringPRL2019}).
A second important point is ensured by the thickness ($120~\mu$m) of the cell : the trapping region must be localized far enough from the walls as to keep fluid parameters constant. The choice of the trapping wavelength (785 nm) also avoids heating locally the fluid. The data presented in the paper are taken from 10 consecutive measurements of 60 seconds each, with an acquisition frequency of $32768$ Hz. The whole experiment is done in constant conditions, with the same bead and only a few seconds between each measurement. This procedure leads to long time-series of $19660800$ positions, spanning over 10 minutes. The concatenation of 10 measurements leads to 10 discontinuities among the $19660800$ points. However, the motion being  confined, these discontinuities are of the same order of magnitude than a regular increment. This together with the small number of such cuts among a large statistics prevent any statistical contribution that would modify the results.

\section{Autocorrelation of displacement}
\label{APPENDIX_cov}

We will compute the autocorrelation function (or covariance, since the process has zero mean) of displacements $dX_t$ defined by the Ornstein-Uhlenbeck process $dX_t = -a X_t dt + b dW_t$ (adopting simple notations $\kappa/\gamma \equiv a$ and $\sqrt{2 k_B T / \gamma}\equiv b $) as:
\begin{equation}
	\begin{aligned}
		\langle dX_t dX_s \rangle &= \langle \left(-a X_t dt + b dW_t\right) \left(-a X_s ds + b dW_s \right) \rangle \\
		&= \underbrace{a^2 \langle X_t X_s \rangle dt^2 }_\text{(1)} -\underbrace{a b \langle X_t dt dW_s \rangle}_\text{(2)} \\
		&-\underbrace{a b \langle X_s ds dW_t \rangle}_\text{(3)} + \underbrace{b^2 \langle dW_t dW_s\rangle}_\text{(4)}  .
	\end{aligned}  \label{autocorr}
\end{equation}
Using the solution of the Ornstein-Uhlenbeck process
\begin{equation}
X_t = X_0 e^{- a t} + b e^{-a t} \int_0^t e^{a t'} dW_{t'},
\end{equation}
and assuming that all time increments are equal ($\forall t, s: dt = ds$), we can compute the different terms in (\ref{autocorr}) one by one:
\begin{equation}
	(1) = \frac{ab^2}{2} e^{-a|t-s|} dt^2
\end{equation}
since at equilibrium $\langle X_0^2 \rangle = k_B T / \kappa = b^2/2a$ (see below Eq. (\ref{APP_Eq:cov})).

\begin{equation*}
	\begin{aligned}
		(2) &=-ab\langle X_t dt dW_s \rangle\\
		 &= -ab\langle X_0 dW_s\rangle e^{-at} dt - ab^2 dt \int_0^t e^{a(t_1 - s)} \langle dW_{t_1} dW_s \rangle\\
		&= -ab \delta(s-0) dt^2 e^{-at} - ab^2 dt \int_0^t e^{a(t_1 - s)} \delta(t_1 - s) ds
	\end{aligned}
\end{equation*}
If we consider non-zero times, we can ignore the first term. For the second, we have two cases :
\begin{equation}
(2) =
  \begin{cases} 
   -ab^2 dt^2 e^{-a(t-s)} & \text{if } t \geq s \\
   0       & \text{if } t < s
  \end{cases}
\end{equation}
Similarly:
\begin{equation}
(3) =
  \begin{cases} 
   0 & \text{if } t > s \\
   -ab^2 dt^2 e^{-a(s-t)} & \text{if } t \leq s
  \end{cases}
\end{equation}
We can therefore combine them into $(2) + (3) = -ab^2 dt^2 e^{-a({\rm max}(t,s) - {\rm min}(t,s)) }$ giving:
\begin{equation}
	(2) + (3) = -ab^2 e^{-a| t-s |} dt^2
\end{equation}
For the forth term, we have simply:
\begin{equation}
	(4) = b^2 \delta(t-s) dt
\end{equation}
that vanishes if $t\neq s$. These 4 terms added together lead to the simple expression of the autocorrelation of displacements:
\begin{equation}
		\langle dX_t dX_s \rangle = - \frac{ab^2}{2} e^{-a|t-s|} dt^2 + b^2\delta(t-s) dt .
\end{equation}
Putting back physical dimensions with $a b^2 dt^2 = \frac{2\kappa k_B T}{\gamma^2} dt^2$ and $\beta^2 dt = \frac{2 k_BT}{\gamma} dt$ (both in $[{\rm m}^2]$), we get
\begin{equation}
		\langle dX_t dX_s \rangle = -\frac{2\kappa k_B T}{\gamma^2} e^{-\kappa|t-s|/ \gamma} dt^2 + 2 D \delta(t-s) dt .
		\label{autocorrdim}
\end{equation}

\begin{figure}[htb]
  \centering{
    \includegraphics[width=1\linewidth]{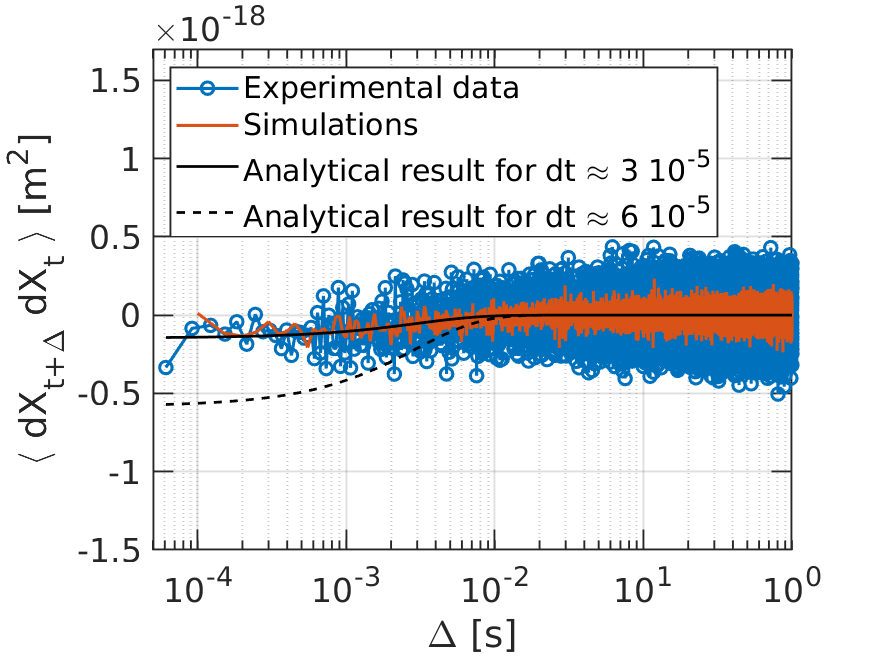}}
  \caption{{Covariance of displacements for the Ornstein-Uhlenbeck process. We plot the experimental result (blue open circles), calculated with $dt = 3.0518\cdot 10^{-5}s$ along with simulation result (orange continuous line) and analytical solution (\ref{autocorrdim}) (black continuous line). We plot (black dashed and continuous lines) the analytical result for two different values of the time-step $dt$ in order to highlight the fact that the deviation from zero of the Ornstein-Uhlenbeck displacements is strongly dependent on the value of $dt$, converging rapidly to zero with increasing acquisition frequency.}}
  \label{APP_fig_cov}
\end{figure}

Since $\langle dW_{t} dW_s \rangle = \overline{ dW_t dW_s }$ for a Wiener process \cite{gardiner}, we can identify the ensemble average $\langle dX_t dX_s \rangle $ with a time averaged covariance $\overline{ dX_t dX_s }$ that is experimentally measured -see Eq.(\ref{cov}) in the main text- and displayed in Fig. \ref{MSD_fig} in the main text and in Fig. \ref{APP_fig_cov} here.

\vspace{3mm}

Fig. \ref{APP_fig_cov} reveals a good agreement between the experimental results, the simulations and the theoretical result (\ref{autocorrdim}). The covariance converges towards zero (which is the covariance of the Wiener increment) for decreasing $dt$. However, the non-differentiability of the stochastic process prevent us from taking the limit of vanishingly small $dt$ and from observing the convergence of the short-time Ornstein-Uhlenbeck process towards a Wiener process.

\section{Derivation of the Mean Square Displacement}
\label{APPENDIX_MSD}

Using the general solution of the Ornstein-Uhlenbeck stochastic differential equation:
\begin{equation}
x_t = x_0 e^{-\kappa t / \gamma} + \sqrt{2D} e^{-\kappa t / \gamma} \int_0^t e^{\kappa t' \gamma} dW_{t'} ,
\end{equation}
we write the expression of the autocorrelation function:
\begin{equation}
\begin{aligned}
	\langle x(t_1) x(t_2) \rangle &= \left(\langle x_0^2\rangle - \frac{k_BT}{\kappa} \right)e^{-\kappa(t_1+ t_2)/\gamma} \\
	&+ \frac{k_BT}{\kappa} e^{-\kappa\mid t_1 - t_2\mid/\gamma}
	\end{aligned}
\end{equation}
that simplifies into:
\begin{equation}
	\langle x(t_1) x(t2) \rangle =\frac{k_BT}{\kappa} e^{-\kappa\mid t_1 - t_2\mid/\gamma}
	\label{APP_Eq:cov}
\end{equation}
if $\langle x_0^2\rangle = \frac{k_BT}{\kappa}$ \textit{i.e.} if the initial distribution is at equilibrium. The MSD therefore writes as:
\begin{equation*}
	 \begin{aligned}
		\langle\delta x^2(\Delta)\rangle &\equiv \langle (x(t+\Delta) - x(t) )^2\rangle \\
		&= \langle x(t+\Delta) ^2\rangle - 2  \langle x(t+\Delta) x(t) \rangle +  \langle x(t) ^2\rangle \\
		&= \frac{k_B T}{\kappa} - 2\frac{k_B T}{\kappa}e^{-\kappa \Delta/\gamma} + \frac{k_B T}{\kappa} , 
	\end{aligned}
\end{equation*}
that is:
\begin{equation}	 
	\langle\delta x^2(\Delta)\rangle = 2 \frac{k_B T}{\kappa}\left(1 - e^{-\kappa \Delta/\gamma}\right) .  \label{MSD_analytic}
\end{equation}
Using the same property of the Wiener process used in Appendix \ref{APPENDIX_cov}, one has $\langle\delta x^2(\Delta)\rangle=\overline{\delta x^2(\Delta) }$ allowing to compare Eq. (\ref{MSD_analytic}) to the experimental result given in Eq. (\ref{msd}) in the main text.

\section{Brownian motion simulations}
\label{APPENDIX_algorithm}

This Appendix briefly presents the structure of the stochastic algorithm, as well as the detailed scheme used for the simulations performed in this article.
The general framework is based on an It\^o-Taylor expansion, generalizing to stochastic differential equations standard Taylor expansion procedures \cite{kloeden}. First, for an ordinary differential equation 
\begin{equation}
	dX_t = a[X_t] dt  ,
\end{equation}
and for a function $f[X_t]$, we can use the standard chain-rule and write $df[X_t] = a[X_t]\frac{\partial }{\partial t}f[X_t]dt $. This leads to an integral form:
\begin{equation}
	f[X_t] = f[X_0] + \int_0^t a[X_s] \frac{\partial f[X_s]}{\partial s} ds
	\label{taylor}
\end{equation}
that can be truncated at a specified order in order to approximate the process described.
In the case of a stochastic process with the following generic form 
\begin{equation}
	dX_t = a[X_t] dt + b[X_t] dW_t  ,
\end{equation}
where $dW_t$ is the stochastic Wiener increment defined by $\langle dW_t \rangle = 0$, $\langle dW_t^2 \rangle = dt $, we use It\^o's lemma instead of the standard chain rule.

We define $\mathcal{L}^0 = a_t \frac{\partial}{\partial X} + \frac{1}{2} b_t^2 \frac{\partial^2}{\partial X^2} $ and $\mathcal{L}^1 = b_t \frac{\partial}{\partial X} $ and use the following notations: $a_t \equiv a[X_t]$ (simil. $b_t$) in order to keep the notation light.
Then similarly to Eq. (\ref{taylor}) we obtain:
\begin{equation}
	f[X_t] = f[X_0] + \int_0^t \mathcal{L}^0f[X_s] ds + \int_0^t \mathcal{L}^1 f[X_s] dW_s .
\end{equation}
We can apply this to $X_t$ itself before iteratively applying it to the quantities $a_t$ and $b_t$ inside the integrals. Doing so, we obtain successive approximations of the process $f[X_t]$ up to a specified order. Hence for the process $X_t$ on a time interval $\Delta t$,  an approximation can be given by:
\begin{equation}
		X_{t+\Delta t} = X_t + a_t \int_t^{t+\Delta t} ds + b_t \int_t^{t+\Delta t} dW_s +  \mathcal{O}(\Delta t^{1}) .
		\label{first_order_trunc}
\end{equation}
By iterating the same procedure up to higher orders, we obtain algorithms with better precision for a given time increment $\Delta t$. The different terms can be written concisely with the following integrals:
\begin{equation}
\begin{aligned}
&\Delta t = \int_{t}^{t +\Delta t} ds\\
&\Delta W = \int_{t}^{t +\Delta t} dW\\
&\Delta Z = \int_{t}^{t +\Delta t} \left[ \int_{t}^{S} dW \right] dS .
\end{aligned}
\end{equation}

With these these definitions, the first order truncation (\ref{first_order_trunc}) gives rise to the Euler-Maruyama scheme for $Y_i$ taken as the numerical approximation of $X_t$:
\begin{equation}
	Y_{i+1} = Y_i + a_i \Delta t + b_i \Delta W_i  ,
\end{equation}
and where the Wiener increment can be simulated by $\Delta W = \eta\sqrt{\Delta t} \equiv \mathcal{N}(0,1) \sqrt{\Delta t} $. Here, the normally distributed random number can be produced by various means, often using built-in functions for random number generation. In our case, the function used is based on the Box-Muller algorithm.

\begin{figure}[htb]
	\centering{
		\includegraphics[width=1\linewidth]{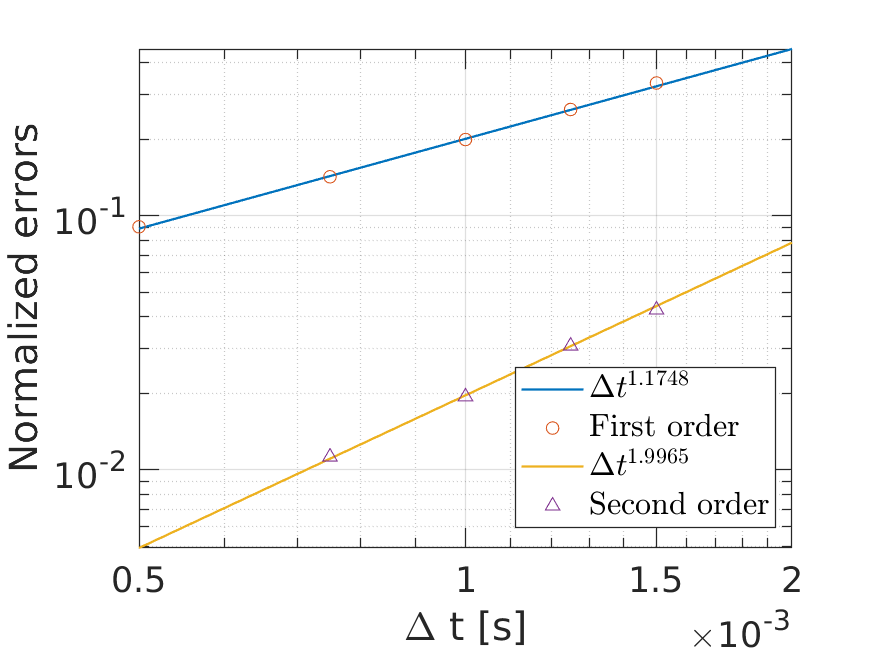}}
	\caption{{Weak convergence test of both Euler-Maruyama and second order algorithms. We plot the errors evaluated as the normalized difference between the measured variance and the theoretical result derived from equipartition $k_B T / \kappa$. Namely $e_{weak} = |1 - \mathbb{E}(Y_i^2)/(k_BT/\kappa)|$ for different values of the time increment $\Delta t$. We observe that the slopes of $\Delta t^{1.1748}$ and $\Delta t^{1.9965}$ are close to the expected ones of $\Delta t^1$ and $\Delta t^2$ respectively.}}
	\label{algo_err}
\end{figure}

In order to evaluate the quality of this algorithm, we rely on the criterion of \textit{weak convergence} \cite{Higham2001}, i.e. convergence of the means. We say that an algorithm has a weak order of convergence $n$ is there exist a constant $C$ such that for all function $f(X_t)$
\begin{equation}
	|\mathbb{E}f(X_t) - \mathbb{E}f(Y_i)| \leq C\Delta t^n .
\end{equation}
In our case we will use $f(X_t) = X_t^2$ and compare the resulting sample variance to its theoretical value. The Euler-Maruyama algorithm is known to converge with weak order $n=1$. We show in Fig. \ref{algo_err} the results of the weak convergence test, giving an exponent $n_{meas.} = 1.1748$.

With the same token, a second order algorithm can be built by keeping the following terms. This gives the following scheme (derived in \cite{kloeden}):
\begin{equation}
\begin{aligned}
	Y_{i+1} &= Y_i + a_i \Delta t + b_i \Delta W_i + \frac{1}{2} b_ib_i' \left( \Delta W_i^2 - \Delta t \right) \\
	& a_i'b_i \Delta Z + \frac{1}{2} \left( a_i a_i' + \frac{1}{2} b_i^2a_i^{(2)} \right)\Delta t^2\\
	& + \left( a_i b_i' + \frac{1}{2}b_i^2b_i^{(2)}\right) \left( \Delta W \Delta t - \Delta Z\right) \\
	& + \frac{1}{2} b_i \left( b_i b_i^{(2)} + (b_i')^2 \right)\left(\frac{1}{3}\Delta W^2 - \Delta t\right).
\end{aligned}
\end{equation}
We can now use the fact that the process we are interested in is defined by $a_t = \kappa X_t / \gamma$ and $b_t = \sqrt{2D}$ which brings all first derivatives of $b_t$ and second derivatives of $a_t$ to zero. With this simplification, we obtain:
\begin{equation}
Y_{i+1} = Y_i + a_i \Delta t + b_i \Delta W_i + b_i a_i' \Delta Z_i + a_i a_i' \Delta t^2 .
\end{equation}
As $\Delta W_i$ is simulated with a random number $\eta$, it is shown in \cite{kloeden} that $\Delta Z$ can be simulated using two independent random numbers $\eta$ and $\theta$, and accordingly: 
\begin{equation}
\small{
	\begin{aligned}
	Y_{i+1} &= Y_i + a_i \Delta t + b_i \sqrt{\Delta t} \eta  \\  
	&+ b_i a_i' \frac{1}{2}\left(\eta + \frac{1}{\sqrt{3}}\theta\right)\Delta t^{3/2} + a_i a_i' \Delta t^2
	\end{aligned}
}
\end{equation}
 This is the weak-$\mathcal{O}(2)$ scheme that we have implemented in a \textsc{Python} code to simulate in the main text the ensembles of Brownian trajectories that are compared to experimental data and to the analytical results. This efficient algorithm reduces numerical errors while keeping a reasonable computing cost.

\vspace{3mm}

\section{Analytical expression of the Allan variance for the Ornstein-Uhlenbeck process (harmonic potential)}
\label{APPENDIX_allan}

For the Ornstein-Uhlenbeck process given by Eq. (\ref{ou}), we have the following power spectral density (PSD) --with $\omega=2\pi f$:
\begin{equation}
S(\omega) = \frac{2 D}{\omega^2 + \omega_0^2},
\end{equation}
where $D = k_B T / \gamma $ is the diffusion coefficient and $\omega_0 = \kappa/\gamma$ corresponds to the trap roll-off frequency. The Allan variance $\sigma^2(\tau) $ is linked to the PSD through a $\sin^4$ transformation, as we discussed previously in \cite{Li}:
\begin{equation}
	\sigma^2(\tau) = \frac{4}{\pi \tau^2} \int_{-\infty}^{+\infty} S(\omega) \sin^4\left(\frac{\omega \tau}{2}\right)d\omega .
\end{equation}
With $\sin^4(x) =\left( e^{4ix} -4e^{2ix} + 6 -4e^{-2ix} + e^{4ix} \right)/16$ and $ \int_{-\infty}^{+\infty} \left( e^{ix} + e^{-ix} \right)dx = 2 \int_{-\infty}^{+\infty}e^{ix}dx$
by parity, we write:
\begin{equation}
	\sigma^2(\tau) = \frac{4}{\pi \tau^2}  \int_{-\infty}^{+\infty} \frac{2D}{\omega^2 + \omega_0^2 }\left( 2e^{2i\omega \tau} - 8e^{i \omega \tau }+ 6 \right)d\omega,
\end{equation}
giving three complex integrals to compute with a simple pole in $\omega = \pm i \omega_0$
\begin{equation*}
\begin{aligned}
	& \int_{-\infty}^{+\infty} \frac{2e^{2i\omega\tau}}{\omega^2 + \omega_0^2} d\omega = 2i\pi {\rm Res}\left( \frac{2e^{2i\omega\tau}}{\omega^2 + \omega_0^2}, i\omega_0 \right) = \frac{2\pi}{\omega_0}e^{-2\omega_0\tau} , \\
	& \int_{-\infty}^{+\infty} \frac{8e^{i\omega\tau}}{\omega^2 + \omega_0^2} d\omega = 2i\pi {\rm Res}\left( \frac{8e^{i\omega\tau}}{\omega^2 + \omega_0^2}, i\omega_0 \right) = \frac{8\pi}{\omega_0}e^{-\omega_0\tau} , \\
	& \int_{-\infty}^{+\infty} \frac{6}{\omega^2 + \omega_0^2} d\omega = 2i\pi {\rm Res}\left( \frac{6}{\omega^2 + \omega_0^2} i\omega_0 \right) = \frac{6\pi}{\omega_0}.
\end{aligned}
\end{equation*}
This done, we obtain
\begin{equation}
\begin{aligned}
	\sigma^2(\tau) &= \frac{8 D}{\pi \tau^2} \frac{1}{16} \left(\frac{2\pi}{\omega_0}e^{-2\omega_0\tau}  -\frac{8\pi}{\omega_0}e^{-\omega_0\tau} +  \frac{6\pi}{\omega_0}\right)\\
	&= \frac{k_B T}{\kappa \tau^2} \left( 4\left[1 - e^{-\kappa\tau/\gamma}\right] - \left[1 - e^{-2\kappa\tau/\gamma}\right]\right)\end{aligned}
\end{equation}
that corresponds to Eq. (\ref{allanvar_eq}) in the main text. Two limits are important to draw:\\
\textit{(i)} the short-time limit $\tau \ll \gamma/\kappa$ where we get $\sigma^2(\tau) \approx 2 D/\tau$ corresponding to free Brownian motion \cite{Li,Oddershede}\\
 \textit{(ii)} the long-time limit $\tau \gg \gamma/\kappa$ where we get a different behavior $\sigma^2(\tau) \approx 3 k_B T / \kappa \tau^2 $.

\section{Analytical expression of the ergodic parameter for the Ornstein-Uhlenbeck process (harmonic potential)}
\label{APPENDIX_ergo}

Under the condition of stationarity, the position correlation function depends only on the time lag $\Delta$ with:
\begin{equation}
C_x(\Delta) =  \left< x(\Delta+t)x(t) \right> = \frac{2k_BT}{\kappa} e^{-\frac{\kappa}{\gamma}\Delta}.  \label{Corr}
\end{equation}
We remind the definition of the ergodic parameter $\epsilon$ \cite{Metzler2014}
\begin{equation}
\epsilon(\Delta) = \frac{\sigma^2\left(\overline{\delta x_i^2(\Delta)}\right)}{\left< \overline{\delta x_i^2(\Delta))} \right>^2} ,
\end{equation}
where $\sigma^2\left(\overline{\delta x_i^2(\Delta)}\right)$ stands for the variance of a single trajectory time average MSD
\begin{equation}
\overline{\delta x_i^2(\Delta)} = \frac{1}{\mathcal{T} - \Delta} \int_0^{\mathcal{T}-\Delta} \left[x_i(t'+\Delta) - x_i(t')\right]^2 dt'  ,
\end{equation}
and $\left< \overline{\delta x_i^2(\Delta))} \right>$ stands for the mean of time average MSD taken over the available ensemble $\{i\}$ of trajectories
\begin{equation}
\left<\overline{\delta x_i^2(\Delta)}\right> = \frac{1}{\mathcal{T} - \Delta} \int_0^{\mathcal{T}-\Delta} \left<\left[x_i(t'+\Delta) - x_i(t')\right]^2\right> dt' .
\end{equation}

Under the ergodic hypothesis, the time ensemble average MSD is: 
\begin{equation}
\left<\overline{\delta x_i^2(\Delta)}\right> = \frac{2k_BT}{\kappa} \left( 1 - e^{-\frac{\kappa}{\gamma}\Delta}\right) ,
\end{equation}
and the variance is defined as:
\begin{equation}
\sigma^2\left(\overline{\delta x_i^2(\Delta)}\right) = \left<\overline{\delta x_i^2(\Delta)}^2\right> - \left<\overline{\delta x_i^2(\Delta)}\right>^2  .
\end{equation}
The first term can be written as
\begin{equation}
\small{
\begin{aligned}
	\left<\overline{\delta x_i^2(\Delta)}^2\right> &= \frac{1}{(\mathcal{T}-\Delta)^2} \int_0^{\mathcal{T} - \Delta} dt_1 \int_0^{\mathcal{T} - \Delta} dt_2 \\
	&\left< (x(t_1+\Delta) - x(t_1))^2(x(t_2 + \Delta) - x(t_2))^2 \right>
\end{aligned}
}  \label{var_allan}
\end{equation}
for which the Wick's relation yields 4 terms:
\begin{equation}
\begin{aligned}
\left<x(t_1) x(t_2) x(t_3) x(t_4)\right> &= \left< x(t_1) x(t_2)\right> \left< x(t_3) x(t_4) \right> \\
&+ \left< x(t_1) x(t_3)\right> \left< x(t_2) x(t_4)\right> \\
&+ \left< x(t_1) x(t_4)\right>\left< x(t_2) x(t_3)\right>  .
\end{aligned}
\end{equation}
The integrand in Eq. (\ref{var_allan}) then becomes:
\begin{equation}
\begin{aligned}
& \left< (x(t_1+\Delta) - x(t_1))^2(x(t_2 + \Delta) - x(t_2))^2 \right>\\
& = [ \left< (x(t_1+\Delta) - x(t_1))^2\right> \left< (x(t_2+\Delta) - x(t_2))^2\right>\\
& + 2\left< (x(t_1+\Delta) - x(t_1))(x(t_2+\Delta) - x(t_2))\right>^2 ] .
\end{aligned}
\label{Wick}
\end{equation}
With the first term in the LHS of Eq. (\ref{Wick}) identified as the square of the time-ensemble average MSD $\left<\overline{\delta x_i^2(\Delta)}\right>^2$, the variance of time average MSD can finally be written as:
\begin{equation}
\small{
\begin{aligned}
\sigma^2(\overline{\delta x_i^2(\Delta)}) &= \frac{2}{(\mathcal{T}-\Delta)^2}  \int_0^{\mathcal{T} - \Delta} dt_1 \int_0^{\mathcal{T} - \Delta} dt_2\\
& \left< (x(t_1+\Delta) -x(t_1))(x(t_2+\Delta) -x(t_2)) \right>^2  \\
&= \frac{2k_B^2T^2}{(\mathcal{T}-\Delta)^2\kappa^2}  \int_0^{\mathcal{T} - \Delta} dt_1 \int_0^{\mathcal{T} - \Delta} dt_2 \\
& \left( 2e^{-\frac{\kappa}{\gamma}\left| t_1-t_2 \right|} - e^{-\frac{\kappa}{\gamma}\left| t_1-t_2 + \Delta\right|} - e^{-\frac{\kappa}{\gamma}\left| t_2-t_1 + \Delta\right|}\right)^2 ,
\end{aligned}
}
\end{equation}
using Eq.(\ref{Corr}). 

\begin{figure}[htb]
  \centering{
    \includegraphics[width=0.8\linewidth]{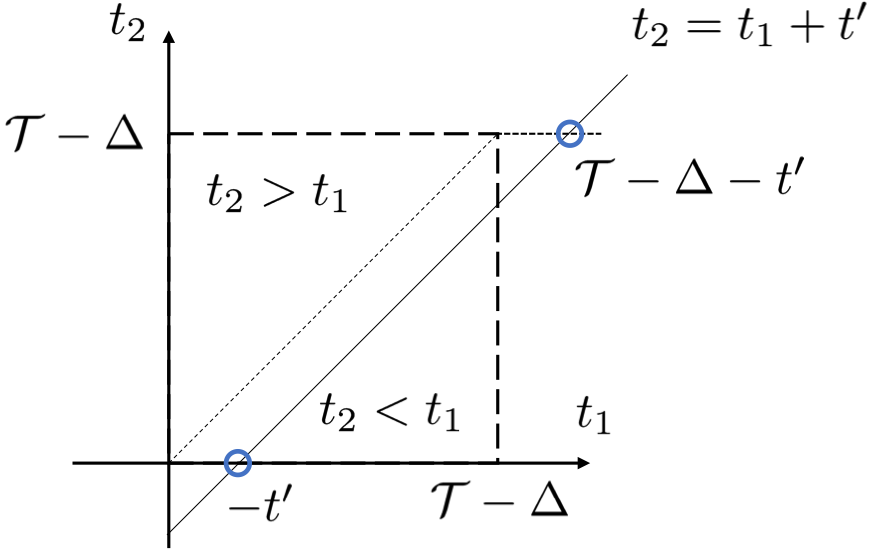}}
  \caption{{Integration surface for Eq. (\ref{intalpha}) on which the two sectors $ [ t_2>t_1 ] $ and $ [ t_2<t_1 ]$ are distinguished. This defines the appropriate change of variables $(t_1,t_2)\leftrightarrow (t_1,t^\prime)$, with the line $t_2=t_1+t^\prime$ crossing the $t_2=0$ axis at $-t^\prime$ and the $t_2=\mathcal{T}-\Delta$ axis at $\mathcal{T}-\Delta-t^\prime$.}}
  \label{figS5}
\end{figure}
\vspace{3mm}

\begin{figure}[htb]
  \centering{
    \includegraphics[width=1\linewidth]{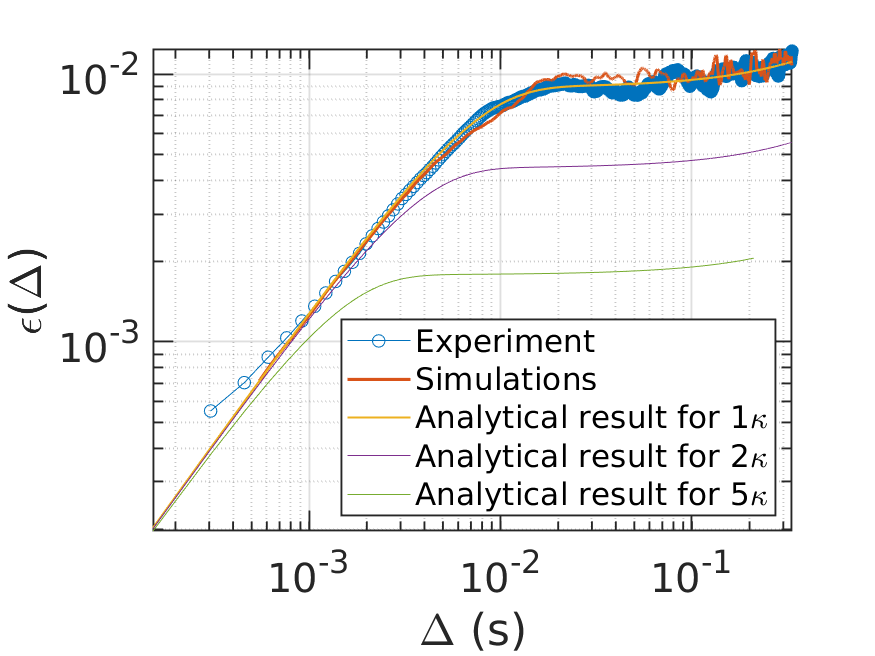}}
  \caption{{Ergodic parameter $\epsilon(\Delta)$ analytically calculated --according to Eq. (\ref{sum_V})-- for 3 different stiffnesses (thin lines). The shift of the plateau and the crossover (roll-off) time clearly appears as $\kappa$ increases. The good agreement between theory and experiment is shown for a stiffness of $1\times\kappa = 2.9614\cdot10^{-6} \pm 6.7339\cdot10^{-8}$ \si{\kilo\gram\per\square\second}.}}
  \label{ergodicity2}
\end{figure}

The integral is calculated through a standard change of variables $t_1=t_1, t' = t_2-t_1$ described in Fig. \ref{figS5} and possible since the integrand only depends on the $| t_1-t_2 |$ difference. One can formally write: 
\begin{equation}
\small{
\sigma^2(\overline{\delta x_i^2(\Delta)}) = \frac{2k_B^2T^2}{(\mathcal{T}-\Delta)^2\kappa^2}  \int_0^{\mathcal{T} - \Delta} dt_1 \int_0^{\mathcal{T} - \Delta} dt_2 \cdot \alpha^2(t') ,
}
\label{intalpha}
\end{equation}
with $t'$ varying from negative to positive values in the $(t_1,t_2)$ plane. For the $t'>0$ sector:
\begin{equation}
\small{
\int_0^{\mathcal{T} - \Delta} dt' \int_0^{\mathcal{T} - \Delta - t'} dt_1 \cdot \alpha^2(t') = \int_0^{\mathcal{T} - \Delta} dt' (\mathcal{T} - \Delta - t') \cdot \alpha^2(t') ,
}
\end{equation}
and for the $t'<0$ sector:
\begin{equation}
\small{
\int_{-(\mathcal{T} - \Delta)}^{0} dt' \int_{-t'}^{\mathcal{T} - \Delta} dt_1 \cdot \alpha^2(t') = \int_{-(\mathcal{T} - \Delta)}^{0} dt' (\mathcal{T} - \Delta + t')  \cdot \alpha^2(t')  .
}
\end{equation}
By combining the two 2 sectors, on gets:
\begin{equation}
\small{
\begin{aligned}
&\int_{-(\mathcal{T} - \Delta)}^{\mathcal{T} - \Delta)} dt' (\mathcal{T} - \Delta -| t' |)  \cdot \alpha^2(t') \\
&= 2 \int_{0}^{\mathcal{T} - \Delta} dt' (\mathcal{T} - \Delta -| t' |)  \cdot \alpha^2(t') \\
&= 2 \int_{0}^{\mathcal{T} - \Delta} dt' (\mathcal{T} - \Delta - t')  \cdot \alpha^2(t')
\end{aligned}
}
\end{equation}
leading to express the ergodic parameter $\epsilon$ as:
\begin{equation}
\small{
\begin{aligned}
\epsilon(\Delta) &= \frac{4k_B^2T^2}{\kappa^2(\mathcal{T}-\Delta)^2\left< \overline{\delta x_i^2(\Delta)} \right>^2}\\
& \int_0^{\mathcal{T}-\Delta} dt' (\mathcal{T}-\Delta-t') \left( 2e^{-\frac{\kappa t'}{\gamma}} - e^{-\frac{\kappa}{\gamma}(t'+\Delta)} - e^{-\frac{\kappa}{\gamma}\left|\Delta - t'\right|} \right)^2
\end{aligned}
}
\end{equation}
In order to simplify the notations, we define $K = \frac{k_BT}{\kappa}$ and write the time ensemble average MSD as
$\left< \overline{\delta x_i^2(\Delta)} \right> = 2K(1-e^{-\frac{\kappa}{\gamma}\Delta})$. The ergodic parameter is then written as $\epsilon(\Delta)=I/4K^2(1-e^{-\frac{\kappa}{\gamma}\Delta})^2$ where the variance of the MSD $I$ is calculated as:
\begin{equation*}
\begin{aligned}
I &= \frac{4K^2}{(\mathcal{T}-\Delta)^2} \int_0^{\mathcal{T}-\Delta} dt' (\mathcal{T}-\Delta-t') \\
&\left( 2e^{-\frac{\kappa t'}{\gamma}} - e^{-\frac{\kappa}{\gamma}(t'+\Delta)} - e^{-\frac{\kappa}{\gamma}\left|\Delta - t'\right|} \right)^2 ,
\end{aligned}
\end{equation*}
splitted in three terms depending on the sign of the absolute value
\begin{widetext}
\begin{equation}
\begin{aligned}
I &= \frac{4K^2}{(\mathcal{T}-\Delta)^2}\mathcal{T} \int_0^{\Delta} \left( 2e^{-\frac{\kappa t'}{\gamma}} - e^{-\frac{\kappa}{\gamma}(t'+\Delta)} - e^{-\frac{\kappa}{\gamma}\left(\Delta - t'\right)} \right)^2dt'  \\
&~~~ -  \frac{4K^2}{(\mathcal{T}-\Delta)^2} \int_0^{\Delta} (t'+\Delta)\left( 2e^{-\frac{\kappa t'}{\gamma}} - e^{-\frac{\kappa}{\gamma}(t'+\Delta)} - e^{-\frac{\kappa}{\gamma}\left(\Delta - t'\right)} \right)^2dt' \\
&~~~ +  \frac{4K^2}{(\mathcal{T}-\Delta)^2} \int_{\Delta}^{\mathcal{T} - \Delta} (\mathcal{T} - \Delta -t')\left( 2e^{-\frac{\kappa t'}{\gamma}} - e^{-\frac{\kappa}{\gamma}(t'+\Delta)} - e^{\frac{\kappa}{\gamma}\left(\Delta - t'\right)} \right)^2dt' \\
&= V_1 + V_2 + V_3 .  \label{sum_V}
\end{aligned}
\end{equation}
Each term is calculated as:
\begin{equation}\nonumber
\begin{aligned}
V_1 = &  \frac{4K^2}{(\mathcal{T}-\Delta)^2}\cdot \frac{\gamma t}{2\kappa} \left[ 5 +\frac{4\kappa}{\gamma}\left(e^{-\frac{2\kappa}{\gamma}\Delta} - 2e^{-\frac{\kappa}{\gamma}\Delta} \right) - 4e^{-\frac{2\kappa}{\gamma}\Delta} + 4e^{-\frac{3\kappa}{\gamma}\Delta} - 4e^{-\frac{\kappa}{\gamma}\Delta} - e^{-\frac{4\kappa}{\gamma}\Delta} \right]  ,  \\
V_2 = &  \frac{K^2}{(\mathcal{T}-\Delta)^2}\cdot \frac{\gamma ^2}{\kappa^2} \Bigg [ \left( 4\frac{\kappa}{\gamma}\Delta - 1 \right) + 12\frac{\kappa^2}{\gamma^2} \Delta^2e^{-2\frac{\kappa}{\gamma}\Delta} \left( 1 - 2e^{\frac{\kappa}{\gamma}\Delta} \right) + \frac{2\kappa}{\gamma} e^{-\frac{4\kappa}{\gamma}\Delta}\Delta \left( -4e^{2\frac{\kappa}{\gamma}\Delta} + 4e^{\frac{\kappa}{\gamma}\Delta} -1 \right)   ,  \\
& + \left( \frac{2\kappa}{\gamma}\Delta +1 \right)\left( -4e^{-2\frac{\kappa}{\gamma}\Delta} + 4e^{-3\frac{\kappa}{\gamma}\Delta} - 4e^{-\frac{\kappa}{\gamma}\Delta} - e^{-4\frac{\kappa}{\gamma}\Delta} +4 \right) + 2e^{-2\frac{\kappa}{\gamma}\Delta} \Bigg ]\\
V_3 = & \frac{16K^2}{(\mathcal{T}-\Delta)^2}\left( \cosh\left( \frac{\kappa}{\gamma}\Delta \right) -1 \right)^2 \Bigg\{ \frac{\gamma}{2\kappa} (\mathcal{T}-\Delta) \left( e^{-2\frac{\kappa}{\gamma}\Delta} -  e^{-2\frac{\kappa}{\gamma}(\mathcal{T}-\Delta)} \right) + \frac{\gamma^2}{4\kappa^2} \bigg [ \left( \frac{2\kappa}{\gamma} (\mathcal{T}-\Delta) +1 \right)e^{-2\frac{\kappa}{\gamma}(\mathcal{T}-\Delta)}\\
& -   \left( \frac{2\kappa}{\gamma} \Delta +1 \right)e^{-2\frac{\kappa}{\gamma}\Delta}  \bigg]  \Bigg\}   ,  
\end{aligned}
\end{equation}
\end{widetext}
whose analytical expression is drawn as the theory curve in Fig. \ref{ergodicity} in the main text.
We show in Fig. \ref{ergodicity2} the impact of the trapping stiffness on the ergodic parameter, clearly displaying how $\kappa$ modifies the long-time plateau as well as the crossover (roll-off) time. We also compare the theory for one specific case with experimental results and numerical simulations.

\vspace{3mm}

\section{Tracking error analysis}
\label{APPENDIX_err}

\subsection{Tracking error on position}

In all our experiments, the trajectories are recorded by a photodiode and the positions are interpreted from the photodiode signal. 
The error on the localization of the particle are originated in our experiments from multiple noise sources dominated by the laser fluctuation and the diode electronic noise. A white noise can be a good starting approximation to estimate and describe the localization error. 
Therefore, each measured position $x_i(t_k)$ for a trajectory $i$ at time $t_k$ can be related to the real position $x_i^0(t_k)$ as \cite{Li,Michalet2012}:
\begin{equation}
x_i(t_k) = x_i^0(t_k) + \mu_i(t_k) ,
\label{tkerr}
\end{equation}
where $\mu_i(t_k)$ is a random uncorrelated tracking error with $\left< \mu_i(t_k) \right> = 0$ and $\left< \mu_i(t_k) \mu_j(t_l) \right> = \delta_{ij} \delta_{kl} \sigma^2_0$. 

\subsection{Tracking error on time ensemble average MSD}
\label{APPENDIX_err_msd}

We now propagate the position tracking error described by Eq.(\ref{tkerr}) into the measured MSD. We write:
\begin{equation}
\begin{aligned}
&\left< \overline{(x_i(t+\Delta) - x_i(t))^2} \right> \\
&= \left< \overline{(x_i^0(t+\Delta) + \mu_i(t+\Delta) - x_i^0(t)- \mu_i(t+\Delta))^2}\right> \\
&= \left< \overline{[(x_i^0(t+\Delta) - x_i^0(t)) +( \mu_i(t+\Delta) - \mu_i(t))]^2}\right>\\
&= \left< \overline{(x_i^0(t+\Delta) - x_i^0(t))^2}\right> + \left< \overline{( \mu_i(t+\Delta) - \mu_i(t))^2} \right>\\
&= \left< \overline{(x_i^0(t+\Delta) - x_i^0(t))^2}\right> + 2\sigma^2_0 ,
\end{aligned}
\end{equation}
showing how the measured MSD can be related to the theoretical one as:
\begin{equation}
\left< \delta x^2(\Delta) \right>_{\rm exp}= \left< \delta x^2(\Delta) \right>_{\rm th} + 2\sigma^2_0 .
\label{MSDerr}
\end{equation}
Since $\sigma_0^2>0$, the MSD is always overdetermined experimentally, in agreement with our observations -in the log-log representation of  Fig. \ref{MSD_fig}, this error can mainly be seen at short time lags. 

\subsection{Tracking error on Allan variance}
\label{APPENDIX_err_allan}

From the definition of Allan variance, we can also relate the experimental Allan variance that includes the tracking errors to the theoretical Allan variance with
\begin{equation}
\small{
\begin{aligned}
\sigma^2_{\rm exp}(\Delta) &= \frac{1}{2\Delta^2} \left< \big(x((n+2)\Delta) - 2x((n+1)\Delta) + x(\Delta)\big)^2 \right>\\
&= \frac{1}{2\Delta^2} \left< \big(x^0((n+2)\Delta) - 2x^0((n+1)\Delta)+ x^0(\Delta)\right.\\
&~~~+\left. \mu_1 - 2\mu_2 + \mu_3 \big)^2 \right>\\
&= \sigma_{\rm th}^2(\Delta) + \frac{1}{2\Delta^2} \left< \big(\mu_1 - 2\mu_2 + \mu_3 \big)^2 \right> \\
&=  \sigma_{\rm th}^2(\Delta) + \frac{3\sigma_0^2}{\Delta^2}  .
\end{aligned}
}
\end{equation}
The difference $3\sigma^2_0/\Delta^2$ between experimental and theoretical Allan variances is always positive and decays with $\Delta^2$, again a feature perfectly consistent with our observations --see Fig. \ref{ergodicity} in the main text.

\begin{figure}[htb]
  \centering{
    \includegraphics[width=0.8\linewidth]{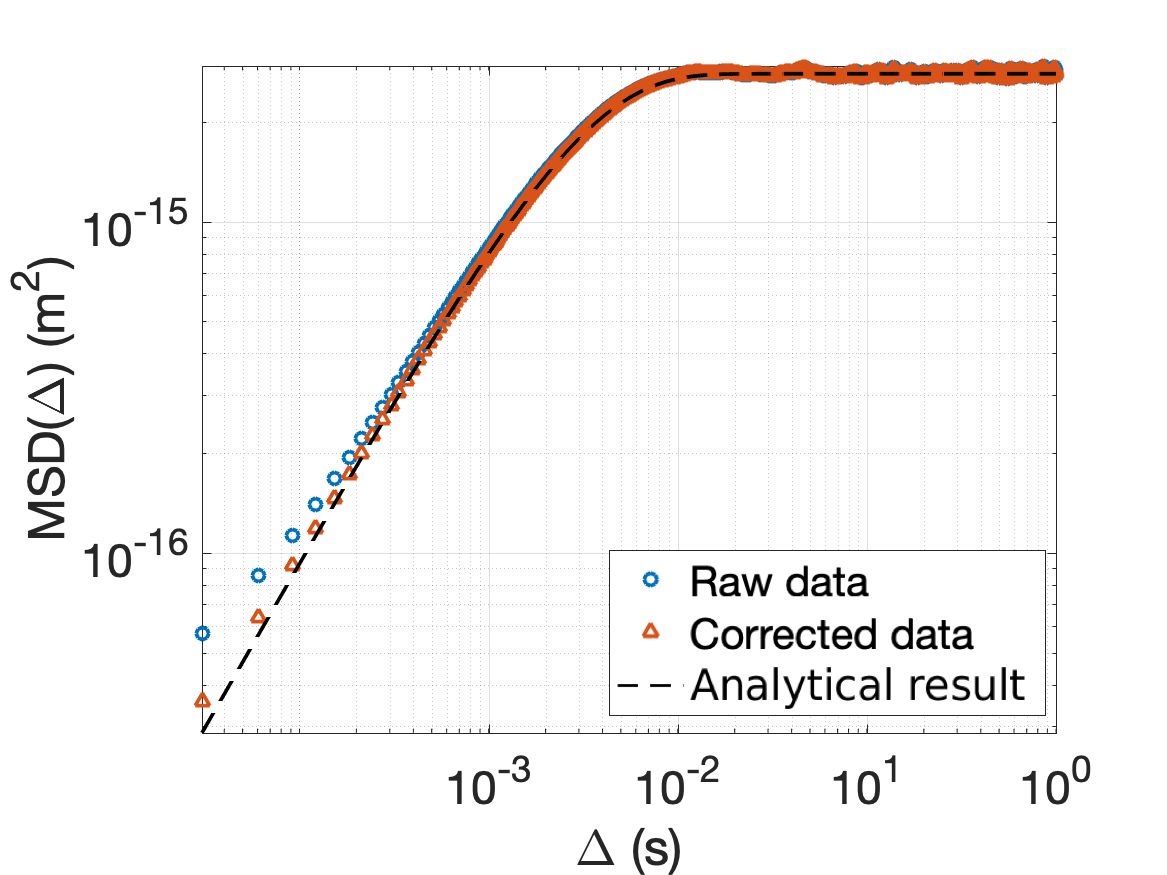}
    \includegraphics[width=0.8\linewidth]{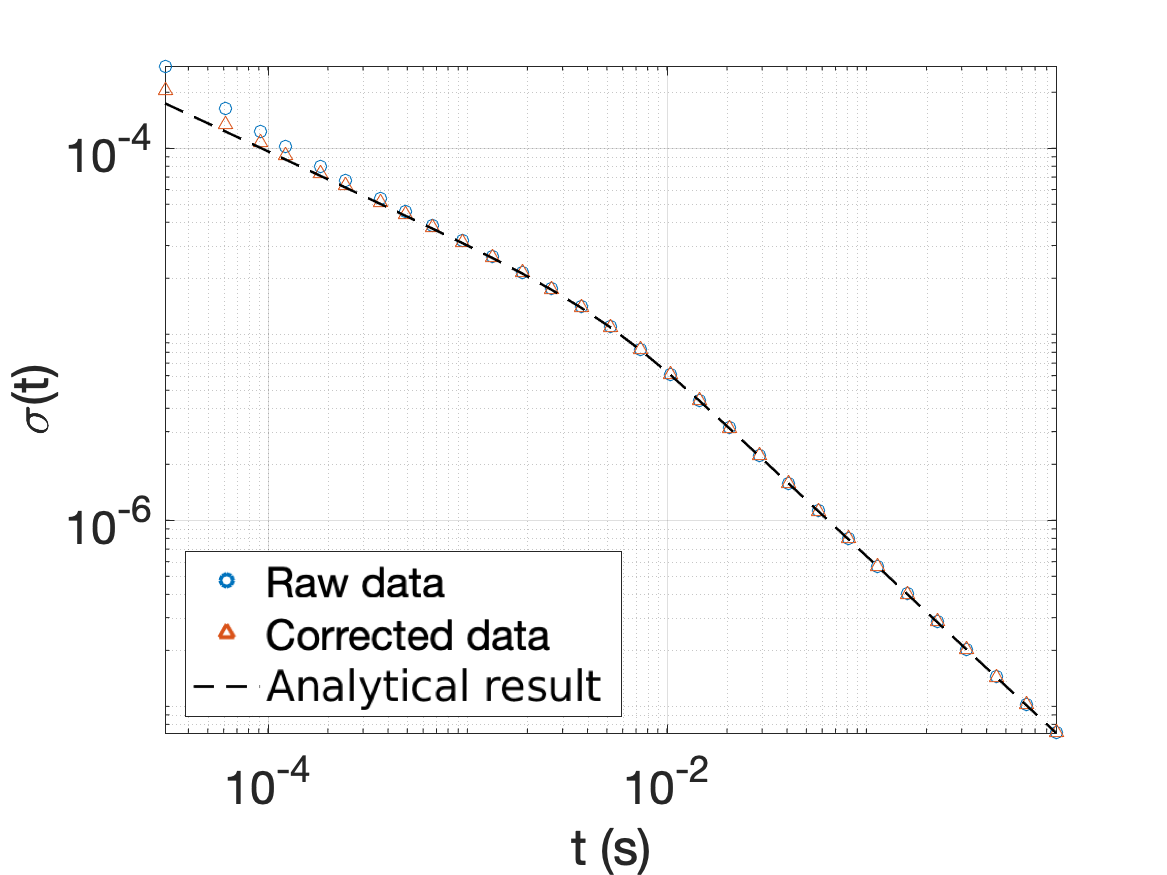}
    \includegraphics[width=0.8\linewidth]{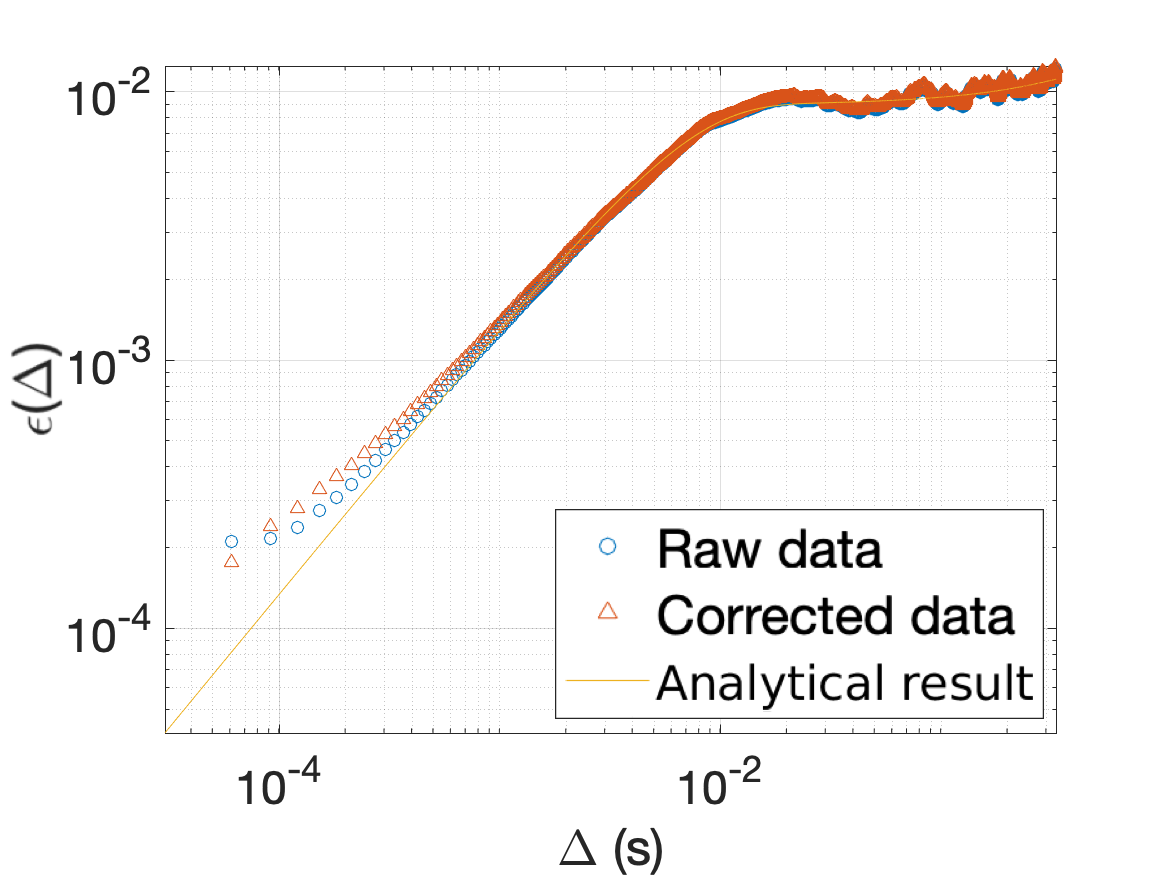}
    }
  \caption{{Raw experimental and corrected data (top) for the MSD, (middle) for the Allan variance and (bottom) for the ergodic parameter. We see the correction mostly for short time-lags. The correction works well for the MSD and Allan variance, but a deviation remains on the ergodic parameter. This difference could actually point to a slight deviation of the localization noise from the white Gaussian noise we have assumed in our modeling of the localization error.}}
  \label{corrections_fig}
\end{figure}

\subsection{Tracking error on the ergodic parameter} \label{APPENDIX_err_erg}

In order to account for the error on the ergodic parameter $\epsilon (\Delta)$, we first consider Eq.(\ref{MSDerr}) for the MSD error analysis. For the single trajectory time averaged MSD, one has
\begin{equation}
\overline{\delta x_i^2(\Delta)}_{\rm exp} = \overline{\delta x_i^2(\Delta)}_{\rm th} + \mu_i
\label{expth}
\end{equation}
where $\mu_i$ is a random constant with $\left< \mu_i \right>^2 = 2\sigma_0^2$. The experimental ergodic parameter can thus be written as:
\begin{equation}
\small{
\epsilon(\Delta)_{\rm exp} = \frac{\left< \left(\overline{\delta x_i^2(\Delta)}_{\rm exp}\right)^2 \right>}{\left<\overline{ \delta x_i^2(\Delta)}\right>^2_{\rm exp}} - 1 = \frac{\left< \left(\overline{\delta x_i^2(\Delta)}_{\rm th} + \mu_i \right)^2 \right>}{\left< \overline{\delta x_i^2(\Delta)}\right>^2_{\rm exp}} - 1  .
} 
\end{equation}
We define the ratio
\begin{equation}
\phi(\Delta) = \frac{\left< \overline{\delta x_i^2(\Delta)}\right>_{\rm th}}{\left< \overline{\delta x_i^2(\Delta)}\right>_{\rm exp}}
\end{equation}
as the ratio between the theoretical and experimental MSD variance value. With this ratio, the experimental ergodic parameter $\epsilon(\Delta)$ can be written as:
\begin{equation}
\small{
\begin{aligned}
\epsilon(\Delta)_{\rm exp} &= \phi^2(\Delta) \epsilon(\Delta)_{\rm th} \\
&+ \phi^2(\Delta) \left( \frac{2\left< \mu_i \overline{\delta x_i^2(\Delta)}_{\rm th}\right> + \left<\mu_i^2\right>}{\left< \overline{\delta x_i^2(\Delta)}\right>^2_{\rm th}} +1 \right) -1 .
\end{aligned}
}
\end{equation}
Assuming that the error $\mu_i$ is uncorrelated with the single trajectory time ensemble MSD, $\langle\epsilon_i \bar{\delta x_i(\Delta)_{th}}\rangle = \langle\epsilon_i\rangle\langle\bar{\delta x_i(\Delta)_{th}}\rangle$. Taking this into account additionally leads to $\left<\mu_i^2\right> = \left<\mu_i\right>^2 + \sigma^2(\mu_i) = 2\sigma_0^2 + \sigma^2(\mu_i)$ and therefore to:
\begin{equation}
\epsilon(\Delta)_{\rm exp} = \phi^2(\Delta)\left[ \epsilon(\Delta)_{\rm th} +  \frac{\sigma^2(\mu_i)}{\left< \overline{\delta x_i^2(\Delta)}\right>^2_{\rm th}}\right] .
\end{equation}
The ratio 
\begin{equation}
\phi(\Delta) = \frac{1}{1+ \frac{\kappa\sigma^2_0}{k_BT(1-e^{-\kappa\Delta/\gamma})}}
\end{equation}
can be estimated once the value of the localization error $\sigma^2_0$ is known. As for the variance of $\mu$, Eq. (\ref{expth}) gives:
\begin{equation}
\sigma^2(\mu_i) = \sigma^2(\overline{\delta x^2_i(\Delta)}_{\rm exp}) -  \sigma^2(\overline{\delta x^2_i(\Delta)}_{\rm th}).
\end{equation}
Since $ \sigma^2(\overline{\delta x^2_i(\Delta)}_{\rm th})$ goes to zero when $\Delta \rightarrow 0$, one is left, at small $\Delta$ with $\sigma^2(\epsilon) \sim \sigma^2(\overline{\delta x^2_i(\Delta)}_{\rm exp})$. Taking the experimental variance measured on the time average MSD for the smallest time lag $\Delta$ is therefore a good estimation for $\sigma^2(\mu)$. This analysis leads us to approaching the real value of the tracking error on the parameter $\epsilon(\Delta)$ and this way explaining the difference between the experimental data and the theoretical curve in Fig. \ref{ergodicity} and \ref{ergodicity2}.


\bibliography{biblio_Goerlich_ms.bib}

\end{document}